\documentstyle[ltwol]{article}

\input{psfig}

\def\Journal#1#2#3#4{{#1} {\bf #2}, #3 (#4)}

\def\PLB{{\em Phys. Lett.}  B}
\def\PRL{\em Phys. Rev. Lett.}
\def\PR{\em Phys. Rep.}
\def\PRD{{\em Phys. Rev.} D}

\def\nn{\noindent}
\def\ie{{\it i.e.}}
\def\eg{{\it e.g.}}

\def\etal{{\it et al.}}
\def\ibid{{\it ibid}.}

\def\be{\begin{equation}}
\def\ee{\end{equation}}
\def\bea{\begin{eqnarray}}
\def\eea{\end{eqnarray}}

\begin{document}

\onecolumn

\vbox{ \large
\begin{flushright}
SLAC-PUB-7926\\
\end{flushright}

\vspace{1in}

\begin{center}
{Resonant Slepton Production at Hadron Colliders in $R$-parity 
Violating Models}
\medskip
\vskip .3cm

{\large J.L. Hewett and T.G. Rizzo }\\
Stanford Linear Accelerator Center, Stanford, CA 94309, USA\\
\vskip 1.0cm

\vskip3cm

{\bf Abstract}\\
\end{center}
Single $s$-channel production of sleptons, such as $\tilde \tau$'s and/or 
$\tilde \nu_\tau$, with their subsequent decay into purely leptonic or dijet 
final states is possible in hadronic collisions via $R$-parity violating 
couplings. We examine the impact of slepton production on bump searches in both 
the Drell-Yan and dijet channels and examine whether the lepton charge 
asymmetry in the $\ell\nu$ channel provides for additional search sensitivity. 
As a consequence, search reaches in the slepton mass-$R$-parity violating 
coupling plane are obtained for both the Tevatron and LHC. The possibility of 
using the leptonic angular distributions and the lepton charge asymmetry to 
distinguish slepton resonances from new gauge bosons is also analyzed.

\vspace*{2.5in}
\noindent 
To appear in the {\it Proceedings of the XXIX International 
Conference on High Energy Physics}, Vancouver, CA, 23-29 July 1998
\vskip 1.3cm

\noindent $^*$Work supported by the U.S. Department of Energy under contract
DE-AC03-76SF00515. 

\thispagestyle{empty}
}
\newpage


\title{{RESONANT SLEPTON PRODUCTION AT HADRON COLLIDERS IN $R$-PARITY 
VIOLATING MODELS}}

\author{ {J. L. HEWETT and T. G. RIZZO}}

\address{Stanford Linear Accelerator Center, Stanford University, Stanford, 
CA 94309, USA\\E-mail: hewett,rizzo@slac.stanford.edu}

\twocolumn[\maketitle\abstracts{
Single $s$-channel production of sleptons, such as $\tilde \tau$'s and/or 
$\tilde \nu_\tau$, with their subsequent decay into purely leptonic or dijet 
final states is possible in hadronic collisions via $R$-parity violating 
couplings. We examine the impact of slepton production on bump searches in both 
the Drell-Yan and dijet channels and examine whether the lepton charge 
asymmetry in the $\ell\nu$ channel provides for additional search sensitivity. 
As a consequence, search reaches in the slepton mass-$R$-parity violating 
coupling plane are obtained for both the Tevatron and LHC. The possibility of 
using the leptonic angular distributions and the lepton charge asymmetry to 
distinguish slepton resonances from new gauge bosons is also analyzed.}]

\section{Introduction}

As is well known, the conventional 
gauge symmetries of the supersymmetric extension of the 
Standard Model(SM) allow for the existence of additional terms in the 
superpotential that violate Baryon($B$) and/or Lepton($L$) number. One quickly 
realizes that simultaneous existence of such terms leads to rapid proton 
decay. These phenomenologically dangerous terms can be written as
\begin{equation}
W_R=\lambda_{ijk}L_iL_jE^c_k+\lambda'_{ijk}L_iQ_jD^c_k+\lambda''_{ijk}
U^c_iD^c_jD^c_k+\epsilon_iL_iH\,, 
\end{equation}
where $i,j,k$ are family indices and symmetry demands that $i<j$ in the terms 
proportional to either the $\lambda$ or $\lambda''$ Yukawa couplings. In the 
MSSM, the imposition of the discrete symmetry of $R$-parity removes by brute 
force all of these `undesirable' 
couplings from the superpotential. However, it easy to construct alternative 
discrete symmetries that allow for the existence of either the $L$- or 
$B$-violating terms~{\cite {herbi}} in $W_R$ (but not both kinds). As far as 
we know there exists no strong 
theoretical reason to favor the MSSM over such $R$-parity violating scenarios. 
Since only $B$- or $L$-violating terms survive when this new symmetry is 
present the proton now remains stable in these models. Consequently, 
various low-energy phenomena then provide the only 
significant constraints~{\cite {constraints}} on the Yukawa couplings 
$\lambda,\lambda'$ and $\lambda''$.

If $R$-parity is violated much of the conventional wisdom associated with the 
MSSM goes by the wayside, \eg, the LSP (now not necessarily a neutralino!) is 
unstable and sparticles may now be produced singly. In particular, it is now 
possible that some sparticles can be produced as $s$-channel resonances, thus 
appearing as bumps in cross sections if kinematically accessible. 
In the case of the two sets of trilinear $L$-violating terms in $W_R$, which we 
consider below, 
an example of such a possibility is production of a $\tilde \nu_\tau$ via 
$d\bar d$ annihilation at the Tevatron or LHC (through $\lambda'$ couplings). 
If this sneutrino decays to, \eg, opposite sign leptons (through the $\lambda$ 
couplings) then an event excess, clustered in mass, will be observed in the 
Drell-Yan channel similar to that expected for a $Z'$. Similarly, the 
corresponding process $u\bar d \to \tilde \tau \to \ell \nu$ may also 
occur through these couplings 
and mimics a $W'$ signature. In addition to these leptonic final 
states, both $\tilde \tau$ and $\tilde \nu$ resonances may decay hadronically 
via the same vertices that produced them, hence leading to potentially 
observable peaks in the dijet invariant mass distribution. 
Thus resonant slepton production, first discussed in Ref. {\cite {bumps}}, 
clearly 
offers a unique way to explore the $R$-parity violating model parameter space. 
It is important to note that $R$-parity violation also allows for other 
SUSY particles, such as $\tilde t$ and/or $\tilde b$, to be exchanged 
in the non-resonant $t,u-$channels and contribute to Drell-Yan events. However, 
it can be easily shown that their influence on cross sections and various 
distributions will be quite small if the low energy constraints on the Yukawa 
couplings are satisfied {\cite {leptos}}. 

\vspace*{-0.5cm}
\nn
\begin{figure}[htbp]
\centerline{
\psfig{figure=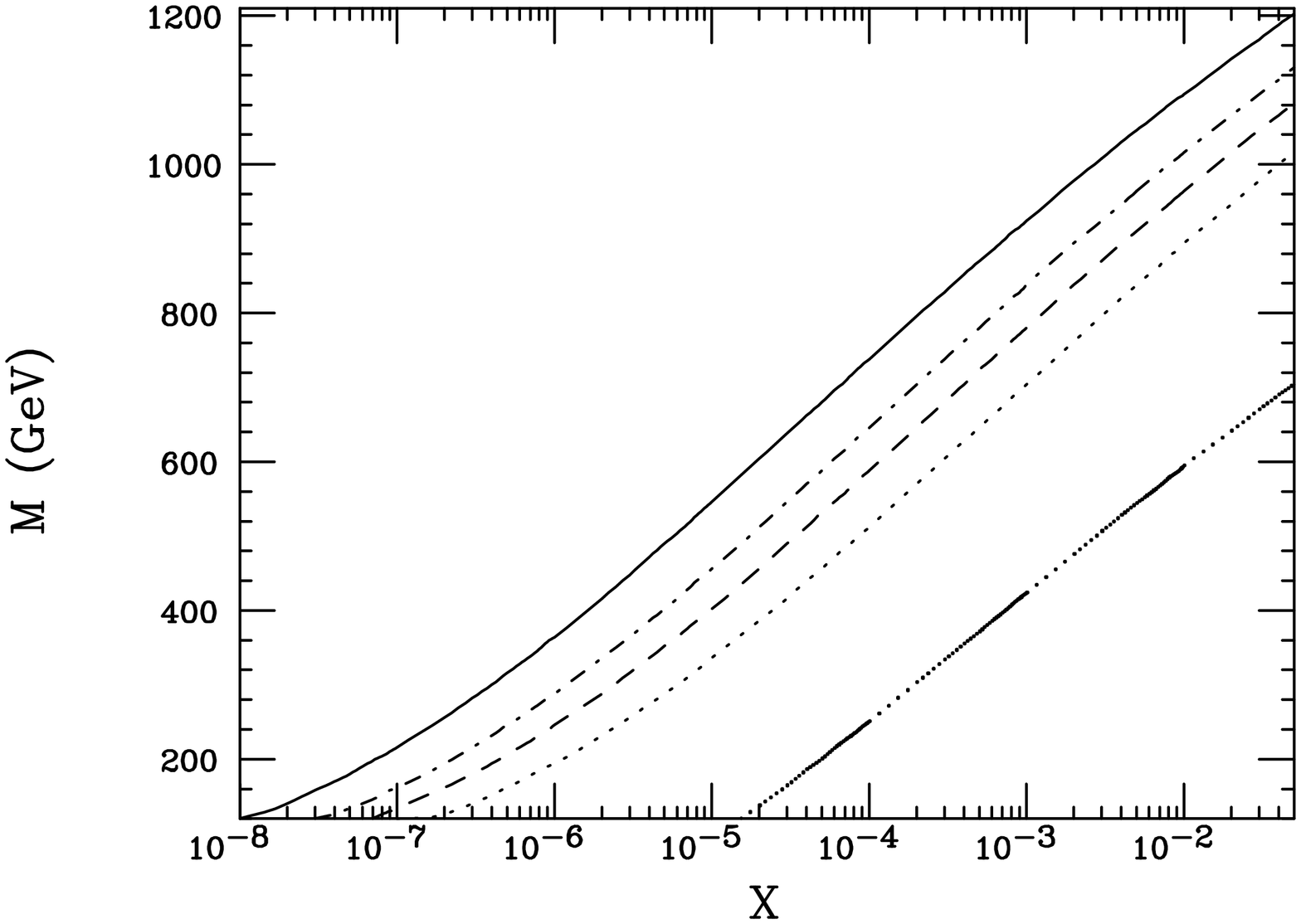,height=4.9cm,width=4.9cm,angle=-90}
\hspace*{-5mm}
\psfig{figure=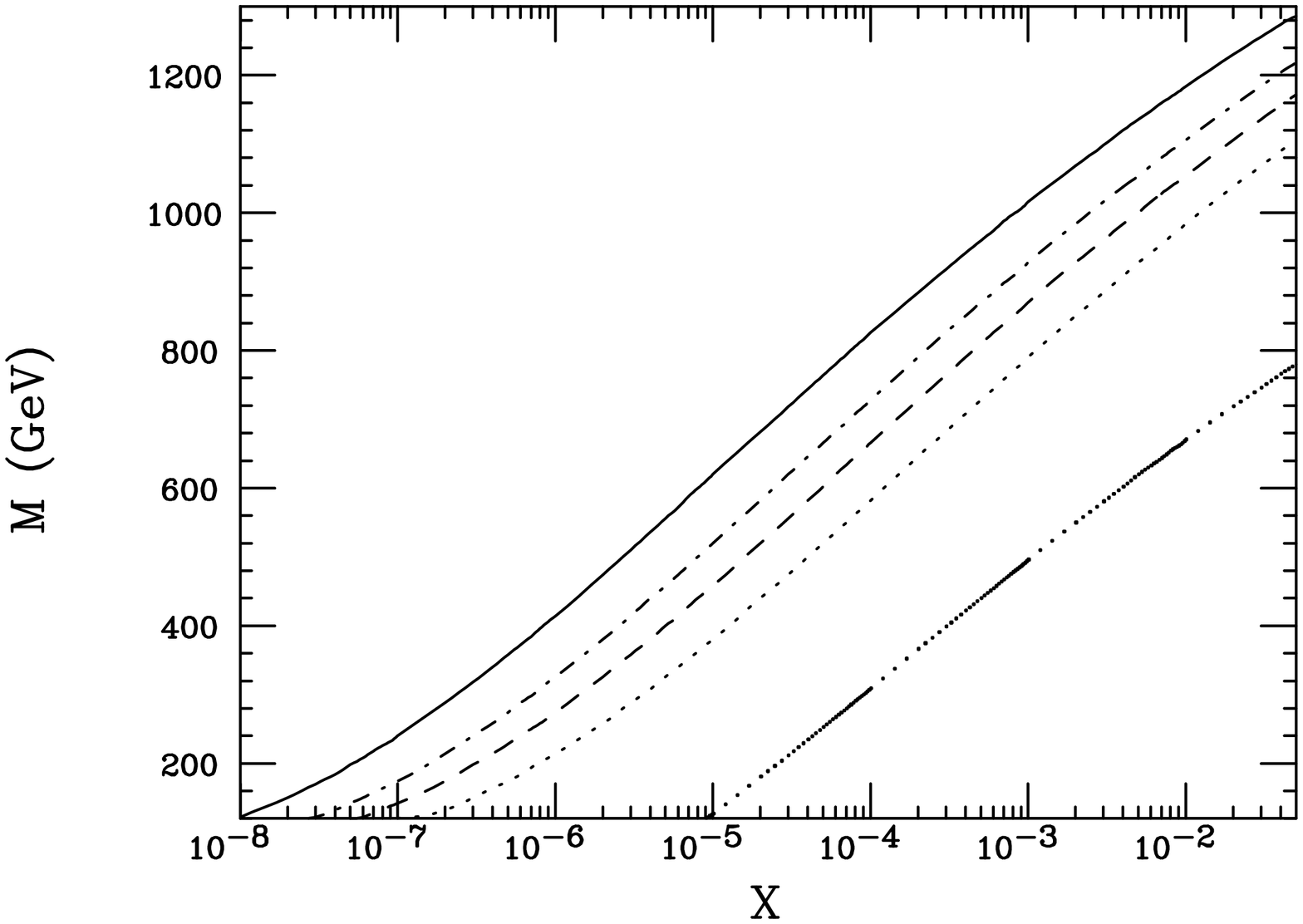,height=4.9cm,width=4.9cm,angle=-90}}
\vspace*{-0.6cm}
\caption{Discovery regions (lying below the curves) in the mass-coupling plane 
for $R$-parity violating resonances in the neutral (left) and charged (right) 
Drell-Yan channels at the Run II Tevatron. From top to bottom the curves 
correspond to integrated luminosities of 30, 10, 5 and 2 $fb^{-1}$. The 
estimated reach for Run I is given by the lowest curve. The parameter X is 
defined in the text.}
\label{tevlim}
\end{figure}
\vspace*{0.1mm}

The questions we address in this analysis are: ($i$) what are the mass and 
coupling reaches for slepton resonance searches at the Tevatron and LHC in 
the Drell-Yan and dijet channels and ($ii)$ how can slepton resonances, once 
discovered, be distinguished from $Z',W'$ production. Below we will mainly 
concern ourselves with the third generation sleptons but our analysis 
is easily extended to those of the first and second generation as well.

\section{The Drell-Yan Channel}

In the case of Drell-Yan production 
the search reach analysis is straightforward being nearly identical to that 
used for new gauge boson production, apart from acceptance issues, \ie, we 
now have spin-0 and not spin-1 resonances. Since sleptons are expected to be 
narrow, the narrow width approximation is adequate and we can directly follow 
the analysis presented in Ref. {\cite {snow}}. In addition to the slepton mass 
itself, the only other parameter in this calculation is the product of the 
appropriate Yukawa couplings, $\lambda'$, 
from the initial state $d\bar u$ or $d\bar d$ 
coupling vertex, and the slepton's leptonic branching fraction, $B_\ell$. 
Calling this product $X=(\lambda')^2B_\ell$,  we can obtain the search 
reach as a function of $X$ in the charged and 
neutral channels for both the Tevatron and LHC; these results are shown in 
Figures 1 and 2. Not only is it important to notice the very 
large mass reach of these colliders for sizeable values of $X\sim 10^{-3}$, 
but we should also observe the small $X$ reach, $X\sim 10^{-(5-7)}$ and 
below, for relatively small slepton masses. Clearly these results show the 
rather 
wide opportunity available to discover slepton resonances over extended 
ranges of 
masses and couplings at these hadron colliders. Note that for fixed values of 
$X$ the search reach is 
greater in the charged current channel due to the higher parton luminosities.

\vspace*{-0.5cm}
\nn
\begin{figure}[htbp]
\centerline{
\psfig{figure=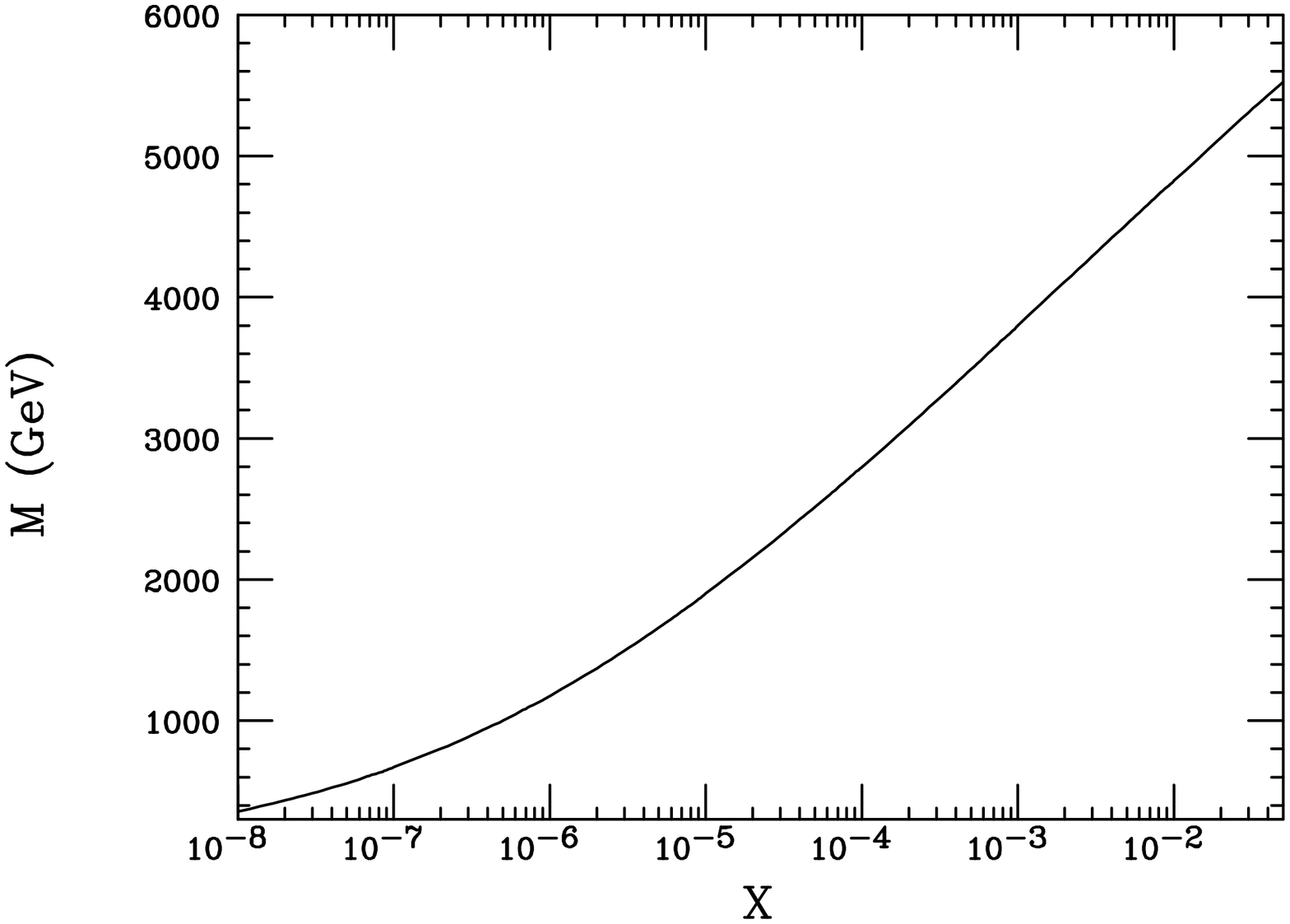,height=4.9cm,width=4.9cm,angle=-90}
\hspace*{-5mm}
\psfig{figure=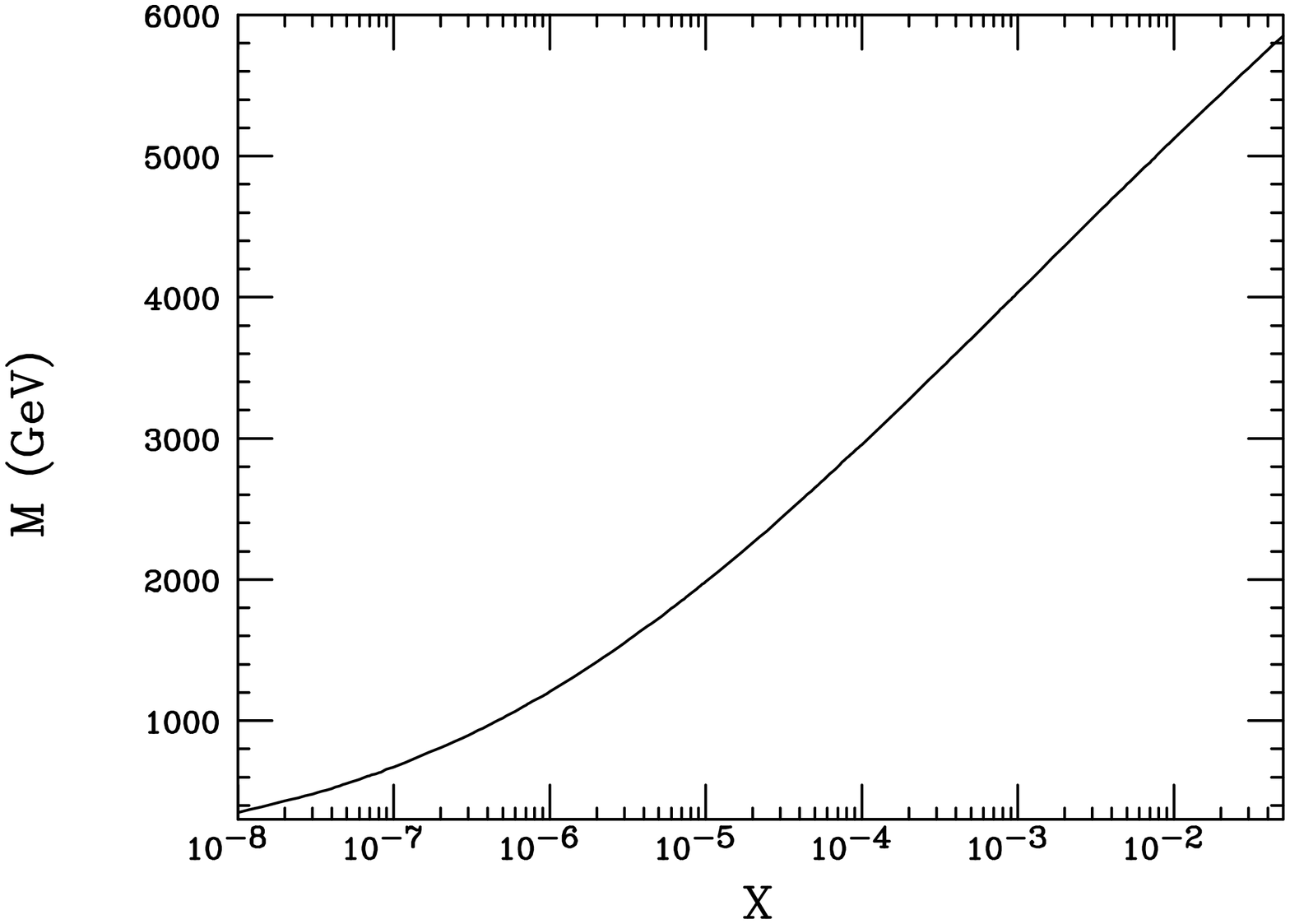,height=4.9cm,width=4.9cm,angle=-90}}
\vspace*{-0.6cm}
\caption{Same as the previous figure but now for the LHC with an integrated 
luminosity of 100 $fb^{-1}$. }
\label{lhclim}
\end{figure}
\vspace*{0.1mm}

Our next issue concerns identifying the resonance (or Jacobian peak) as a 
slepton instead of a new gauge boson. One immediate difference which 
would signal $\tilde \nu$ production would be the observation of the 
very unusual 
$e\mu$ final states which are allowed by the generational structure of the 
superpotential; such final states are not 
expected to occur for a $Z'$ and would be a truly remarkable signature for 
$R$-parity violation. Clearly if the $e\mu$ or SUSY decay 
modes of the slepton dominate there will be no identification problem. If the 
$R$-parity violating modes dominate it is best to look for universality 
violations, \eg, if the 
resonance decays to only one of $e^+e^-$ or $\mu^+\mu^-$ 
or if these two rates are substantially different. Most new gauge bosons which 
are kinematically accessibly are not anticipated to have substantially 
different couplings to the first two fermion generations. In the case of a 
$\tilde \nu$ versus a $Z'$, it is well known 
that most $Z'$ bosons have parity violating 
fermionic couplings which would lead to a forward-backward asymmetry, $A_{FB}$, 
in their leptonic decay distributions. The $\tilde \nu$, being spin-0, would 
always produce a null asymmetry. $A_{FB}$ is more easily measured and 
requires less statistical power than does the reconstruction of the complete 
angular distribution. This is important since, whereas only 10 or so background 
free events would constitute a discovery many more, $\sim 100-200$ are 
required to determine the asymmetry. This would imply that the reach for 
performing this test is somewhat if not substantially less than the 
discovery reach. For example, 
the Tevatron may discovery a $\tilde \nu$ with a mass of 700 GeV for a 
certain value of $X$ but only for masses below 500 GeV would there be enough 
statistics to extract $A_{FB}$ for this same $X$ value.

\vspace*{-0.5cm}
\nn
\begin{figure}[htbp]
\centerline{
\psfig{figure=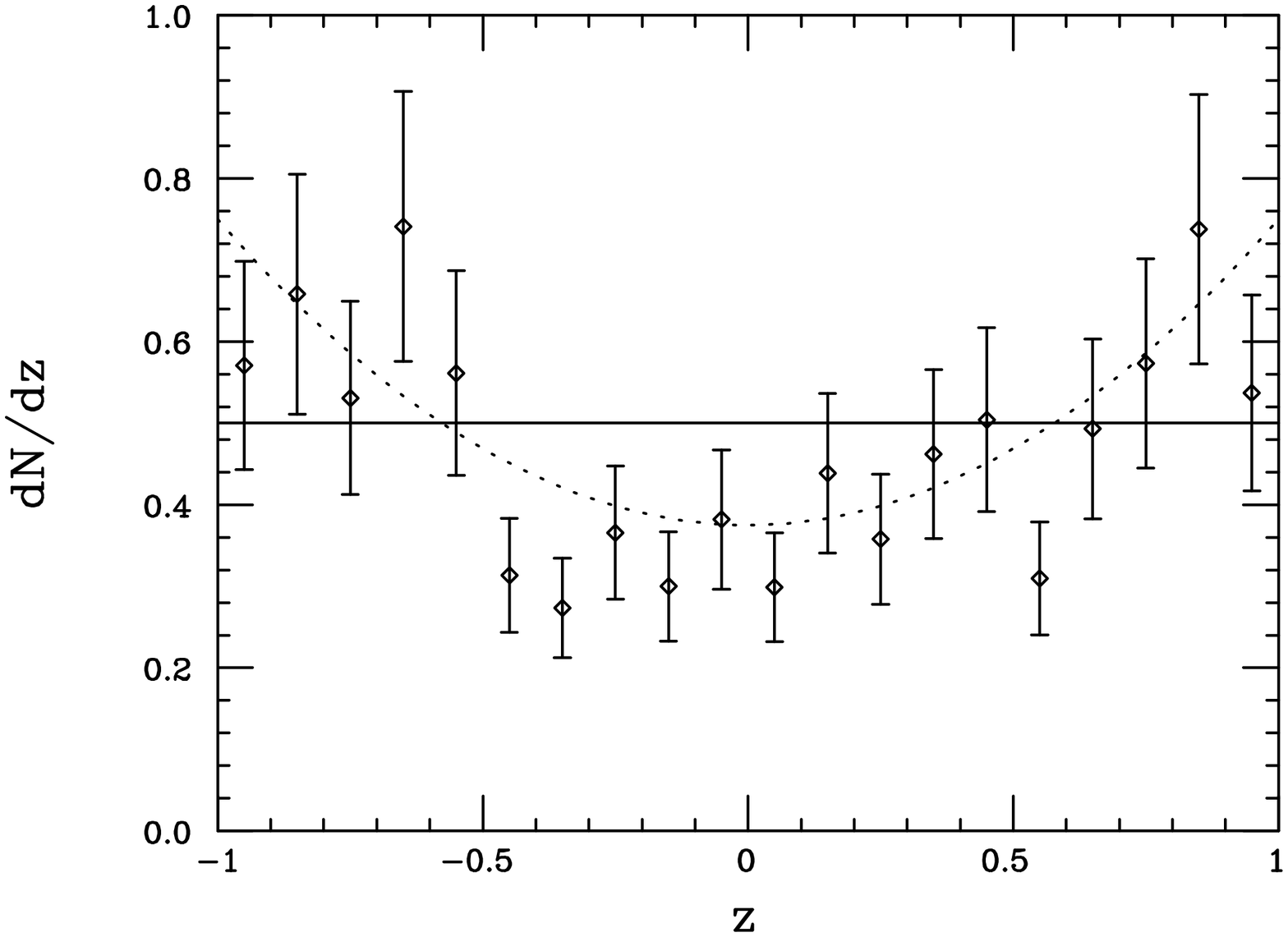,height=4.9cm,width=4.9cm,angle=-90}
\hspace*{-5mm}
\psfig{figure=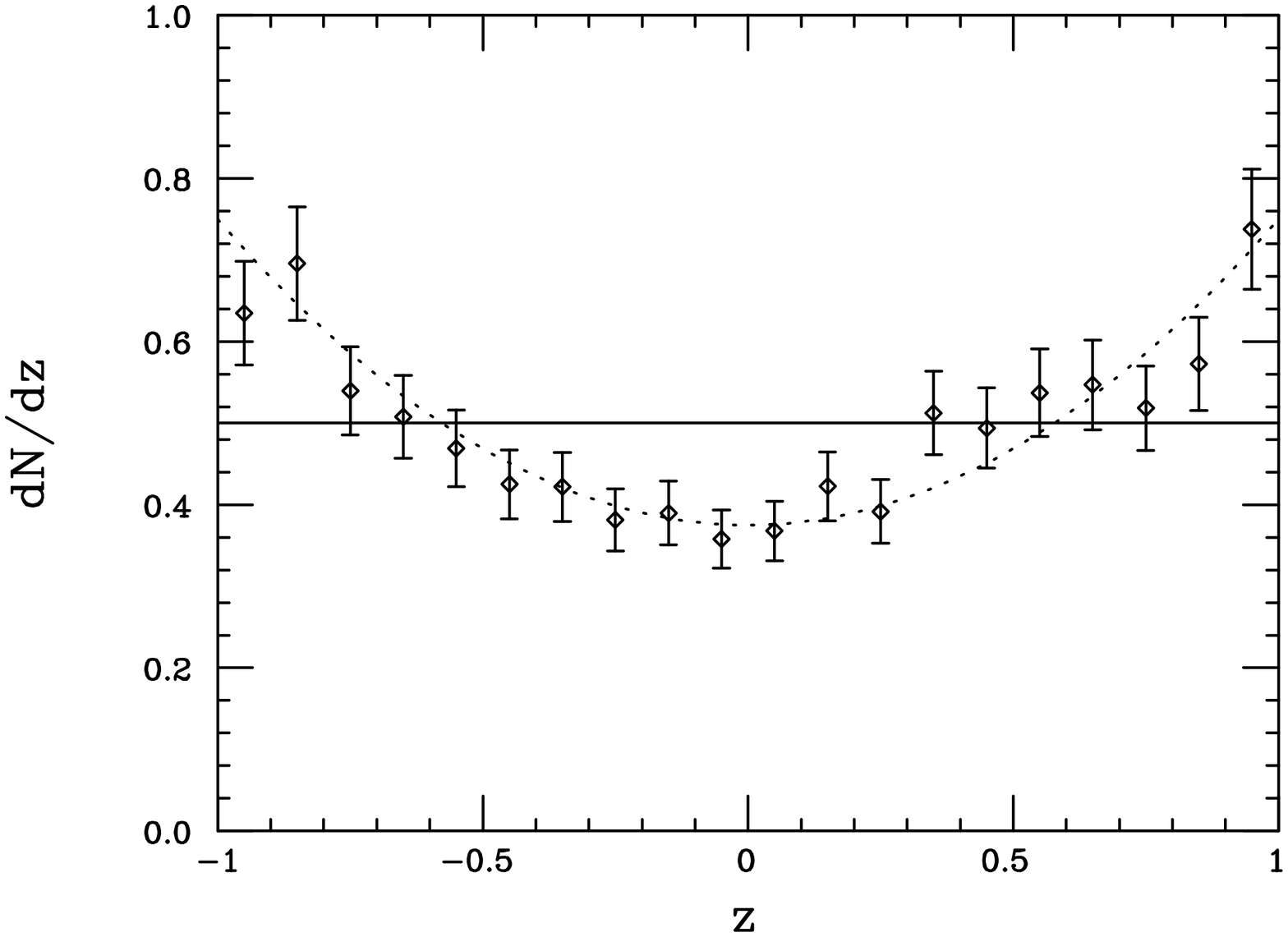,height=4.9cm,width=4.9cm,angle=-90}}
\vspace*{-0.6cm}
\caption{Comparison of the Monte Carlo generated normalized angular 
distribution for the 
leptons in $Z'$ decay with that for a $\tilde \nu$ assuming a (left)400 event 
sample or a (right)2000 event sample; the errors are purely statistical. The 
$Z'$ is assumed to have no 
forward-backward asymmetry due to its fermionic couplings. The $1+z^2$ 
angular distribution is also shown. The effect of potential acceptance losses 
in the two outer bins has not been included.}
\label{angdis}
\end{figure}
\vspace*{0.1mm}

A more complex and interesting situation arises when the $Z'$ naturally has 
$A_{FB}=0$ as in, \eg, some $E_6$ models; {\cite {us}} in this case the 
on-resonance asymmetry data alone is insufficient. If $A_{FB}$ could be 
measured throughout the resonance region, it would be possible to deduce 
through detailed line-shape studies whether or not the new contribution 
interferes with the SM amplitude (something that 
does not occur in the case of $\tilde \nu$ production). Besides requiring 
substantial statistics, finite dilepton mass resolution, especially for the 
$\mu^+\mu^-$ final state, may disrupt this program. 

Of course, with a 
plethora of statistics the complete angular distribution can be obtained as 
shown in Fig.3. Here we compare Monte Carlo generated data for a $Z'$ with a 
zero forward-backward asymmetry with both the flat 
distribution and the $\sim 1+z^2$ distribution hypotheses($z=\cos \theta$) and 
ignore complications due to possible acceptance losses arising from 
rapidity cuts 
in the forward and backward directions. Such a distribution has been measured 
by CDF both on the $Z$ and above. {\cite {angular}}  Both analyses would seem 
to indicate that of order $\sim 1000$ events are required to make a clean 
measurement, a sample approximately 100 times larger than that required for 
discovery. Although such measurements would be conclusive as to the identity 
of the resonance, the required statistics results 
in a significant loss in the mass range over which it can be performed. In 
our Tevatron example above where the search reach was 700 GeV we would find 
that the angular distribution could only be determined for masses below 
$\sim 400$ GeV assuming the same $X$ value.

\vspace*{-0.5cm}
\nn
\begin{figure}[htbp]
\centerline{
\psfig{figure=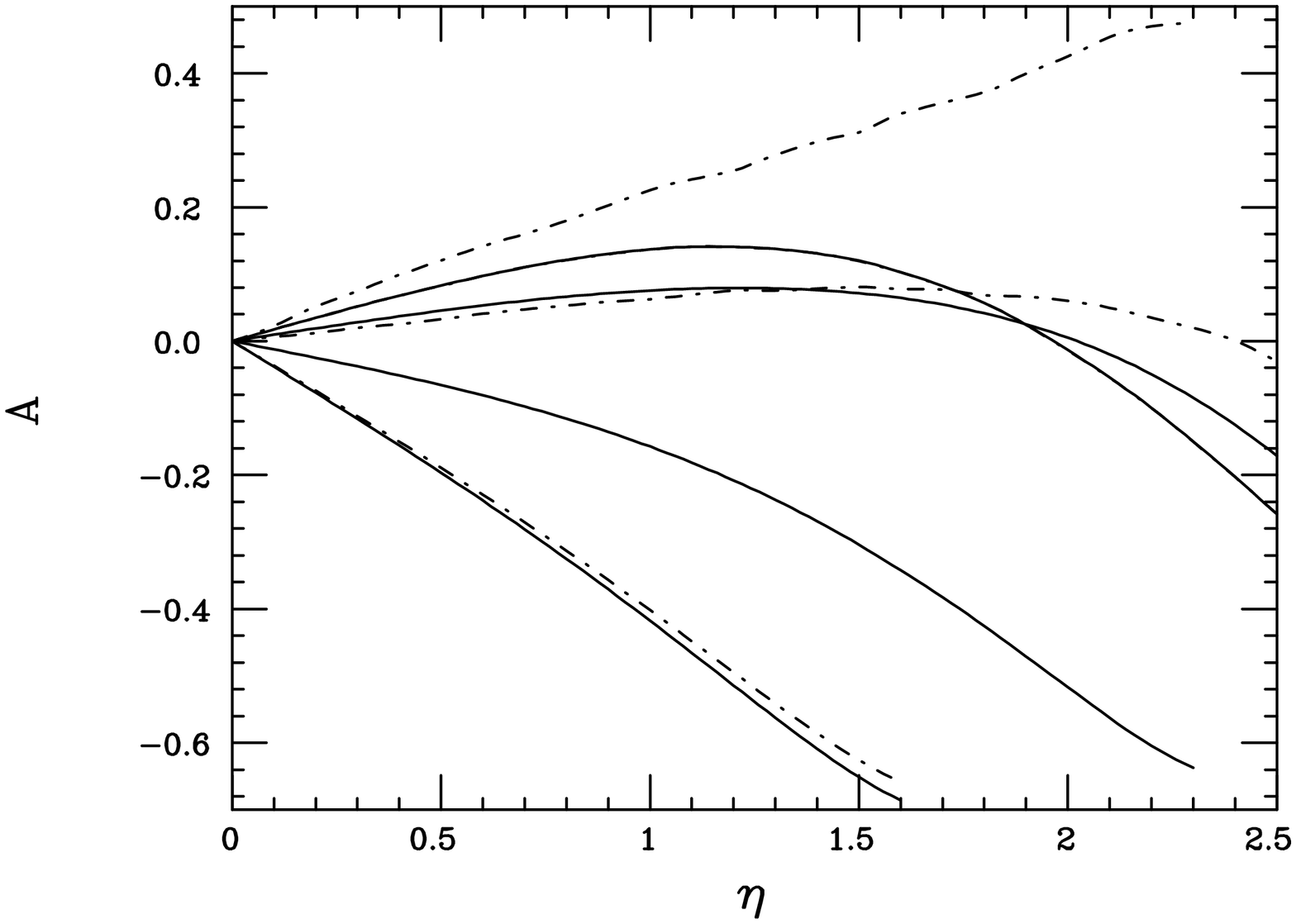,height=4.9cm,width=4.9cm,angle=-90}
\hspace*{-5mm}
\psfig{figure=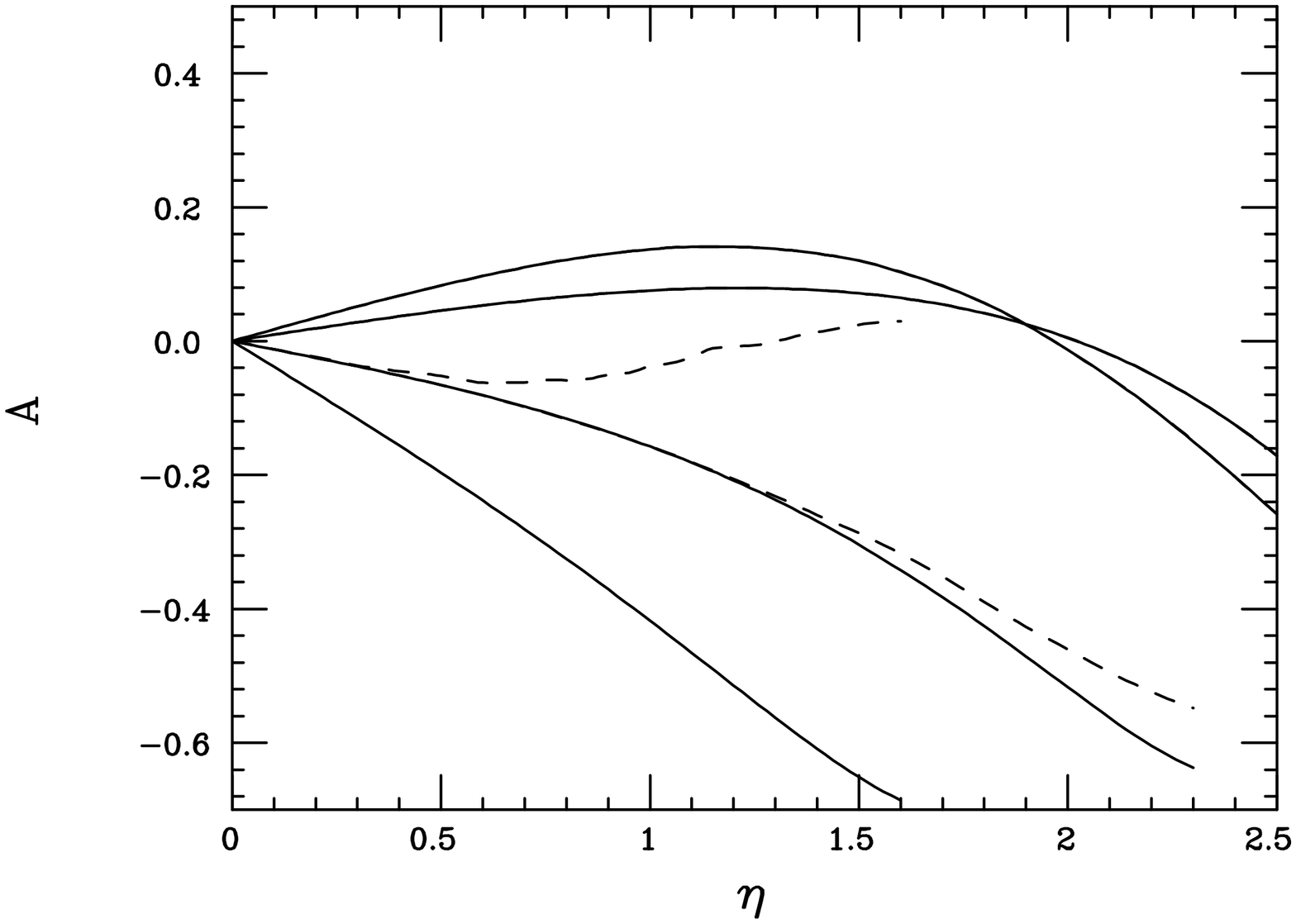,height=4.9cm,width=4.9cm,angle=-90}}
\vspace*{-0.6cm}
\caption{The lepton charge asymmetry in the charged current Drell-Yan 
production channel at the 2 TeV Tevatron for the SM (solid curves)and 
with 250(700) GeV $\tilde \tau$ exchange (the dash-dotted or dashed curves) 
assuming $\lambda,\lambda'=0.15$ for purposes of demonstration. From top to bottom 
in the center of the 
figure, the SM curves correspond to $M_T$ bins of 50-100, 100-200, 200-400 and 
$>400$ GeV, respectively. Note that for $M_T$ in the 50-100 GeV range there 
is no distinction between the SM result and that with a $\tilde \tau$.}
\label{deviations}
\end{figure}
\vspace*{0.1mm}

The angular distribution approach cannot be used to separate between 
the $\tilde \tau$ and $W'$ cases due to the missing energy in the event. 
However, there are two useful observables in this situation.  First, one 
can examine the transverse mass ($M_T$) distribution 
associated with the new Jacobian peak region to see if interference with SM 
amplitudes is occurring. This is far more difficult than in the $\tilde \nu$ 
case again due to the missing energy and mass smearing. 
A second possibility is to examine the leptonic charge asymmetry, 
$A(\eta)$, for the electrons or muons in the final state as a 
function of their rapidity. We remind the reader that $A(\eta)$ is defined as 
\begin{equation}
A(\eta)={dN_+/d\eta -dN_-/d\eta \over {dN_+/d\eta +dN_-/d\eta}}\,,
\end{equation}
where $N_{\pm}$ are the number of positively/negatively charged electrons of a 
given rapidity, $\eta$.  In the SM, the charge asymmetry is sensitive to 
the ratio of u-quark to d-quark parton densities and the $V-A$ production and 
decay of the 
$W$. \cite{chargew}  Since the coupling structure of the $W$ has been 
well-measured, any deviations in this asymmetry within the $M_T$ bin 
surrounding the $W$ have been attributed to modifications in the parton 
distributions (PDF's). Here, we 
are more interested in events with larger $M_T$. Note that $A(-\eta)=-A(\eta)$ 
if $CP$ is conserved(which we assume) so that we will only need to deal with 
$\eta \geq 0$ in the following discussion.

\vspace*{-0.5cm}
\nn
\begin{figure}[htbp]
\centerline{
\psfig{figure=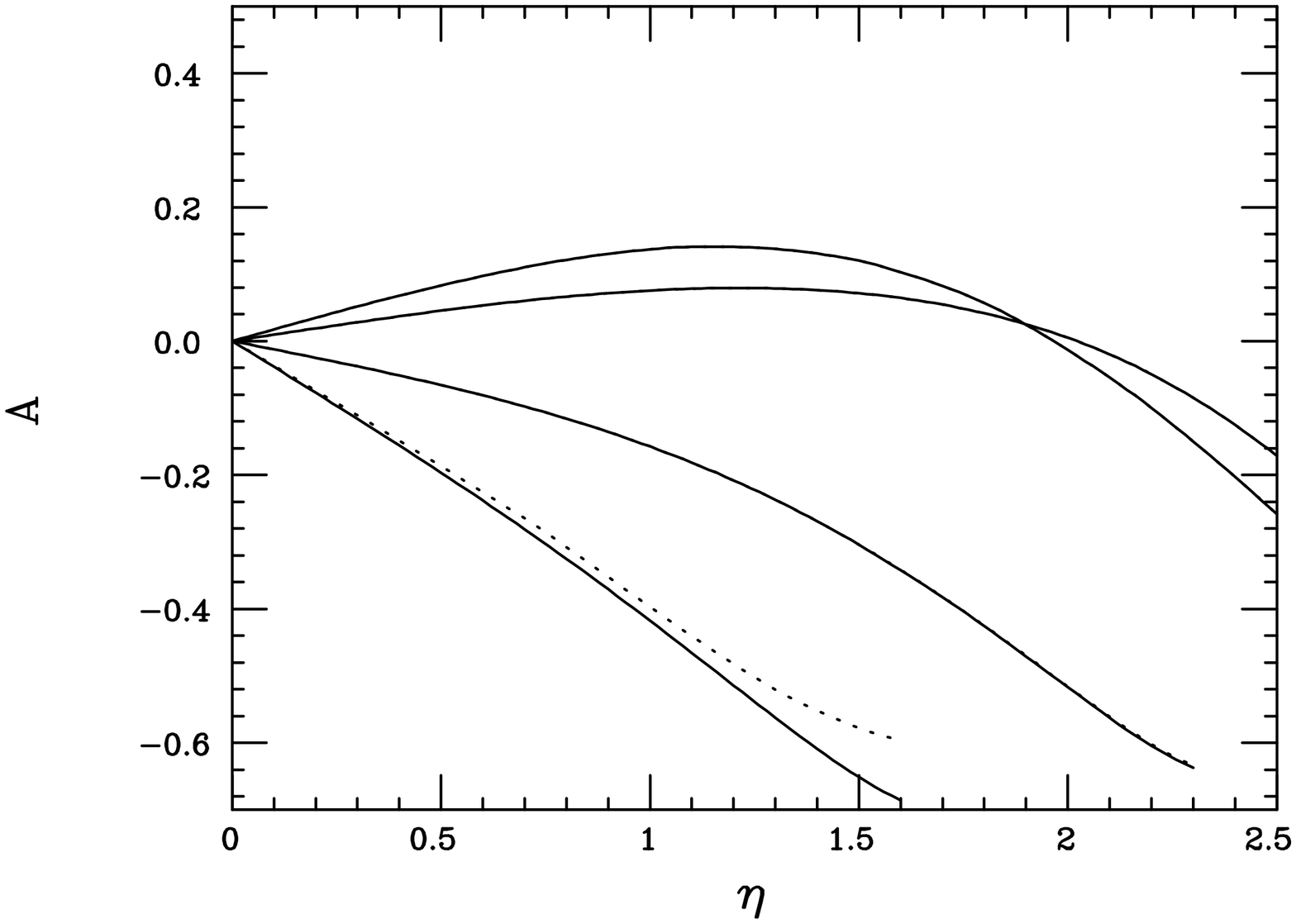,height=4.9cm,width=4.9cm,angle=-90}
\hspace*{-5mm}
\psfig{figure=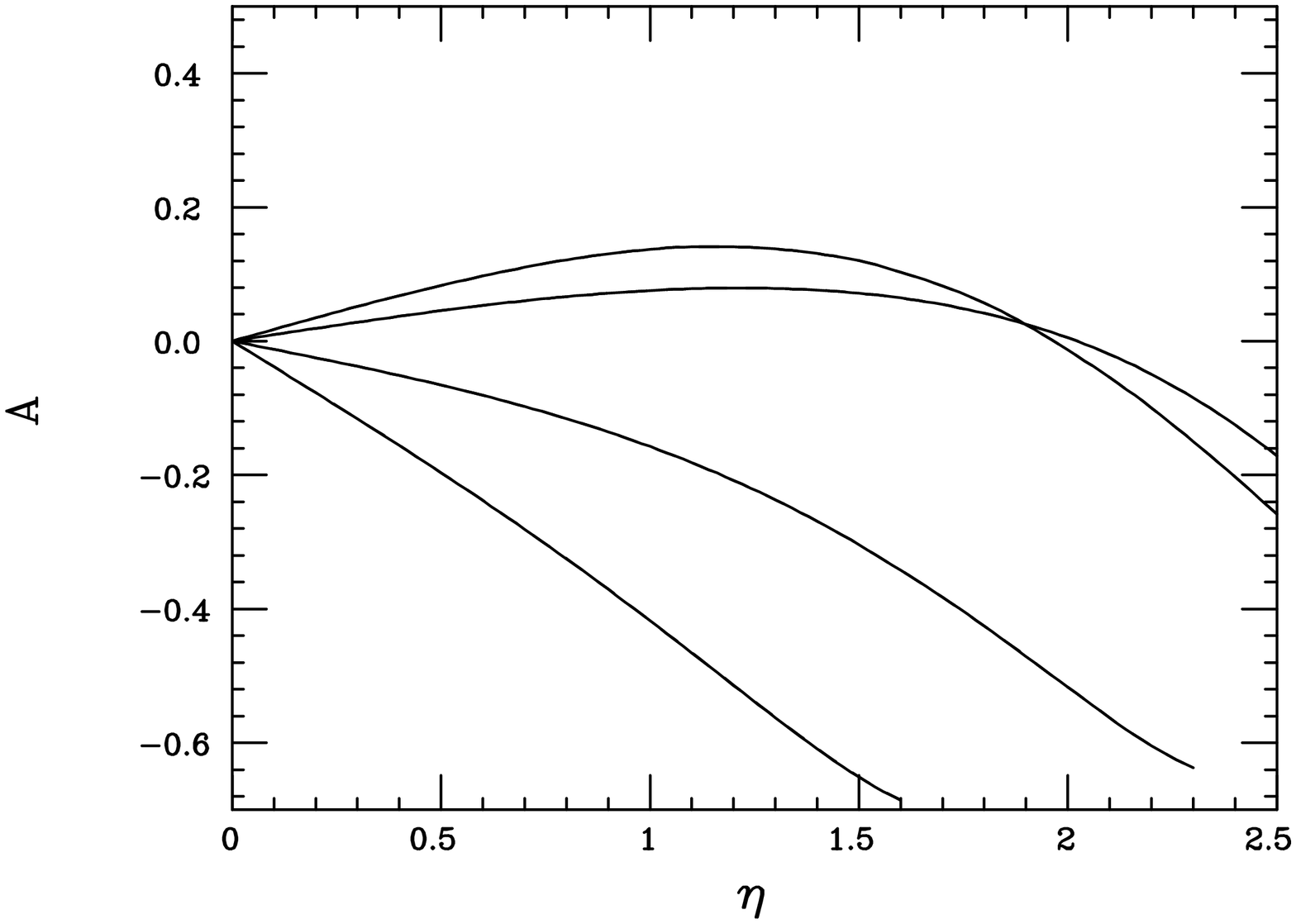,height=4.9cm,width=4.9cm,angle=-90}}
\vspace*{-0.6cm}
\caption{Same as the previous figure but now for 
1000(1500) GeV $\tilde \tau$ exchange corresponding to the the dotted or 
dashed curves in the left(right) panel.}
\label{deviations2}
\end{figure}
\vspace*{0.1mm}

Consider the case for $\tilde \tau$ production at the Tevatron. 
Fig.\ref{deviations} shows the lepton charge asymmetry, within four $M_T$ bins 
corresponding to $50<M_T<100$ GeV, $100<M_T<200$ GeV, $200<M_T<400$ GeV, and 
$400<M_T<1800$ GeV for the SM and how it is modified by the presence of a 
250(700) GeV $\tilde \tau$ with, for 
purposes of demonstration, $\lambda,\lambda'=0.15$.  In particular we observe 
that the lepton 
charge asymmetry can be significantly altered for larger values of $M_T$ in 
the bins associated with the new Jacobian peak.  Note, however, that there 
is essentially no deviation
in the asymmetry in the transverse mass bin associated with the $W$ peak,
$50<M_T<100$ GeV, so that this $M_T$ region can still be used for determination
of the PDFs. Fig.4 also shows that the presence of the $\tilde \tau$ tends to 
drive the asymmetry to smaller absolute values as perhaps might be expected 
due to the presence of a spin-0 resonance which provides a null `raw' 
asymmetry. In Fig.5 we see that the asymmetry is still visible in the last 
$M_T$ bin for the case of a 1 TeV $\tilde \tau$ but becomes essentially 
non-existent for these values of the Yukawa couplings when the mass is raised 
to 1.5 TeV.

\vspace*{-0.5cm}
\nn
\begin{figure}[htbp]
\centerline{
\psfig{figure=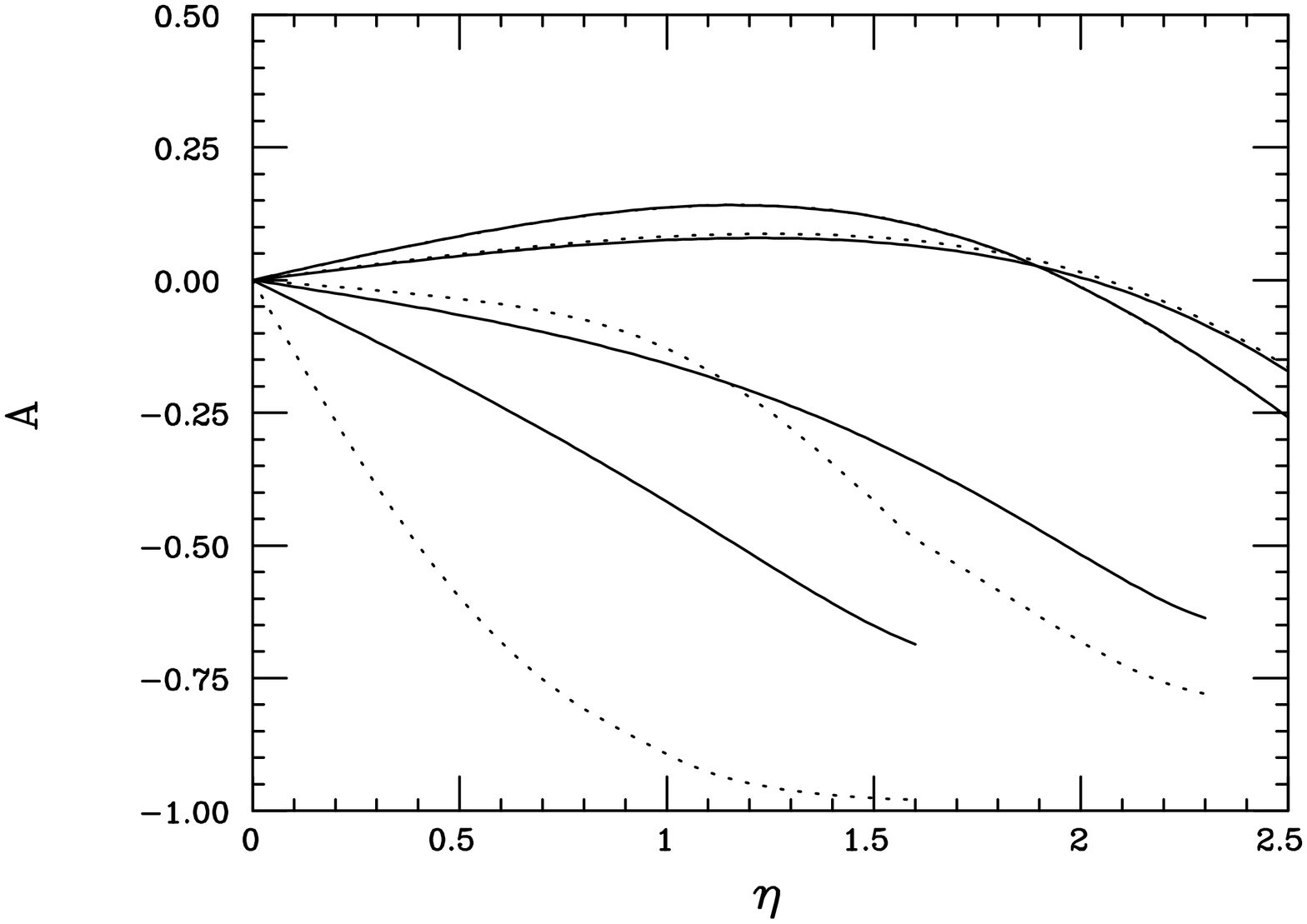,height=4.9cm,width=4.9cm,angle=-90}
\hspace*{-5mm}
\psfig{figure=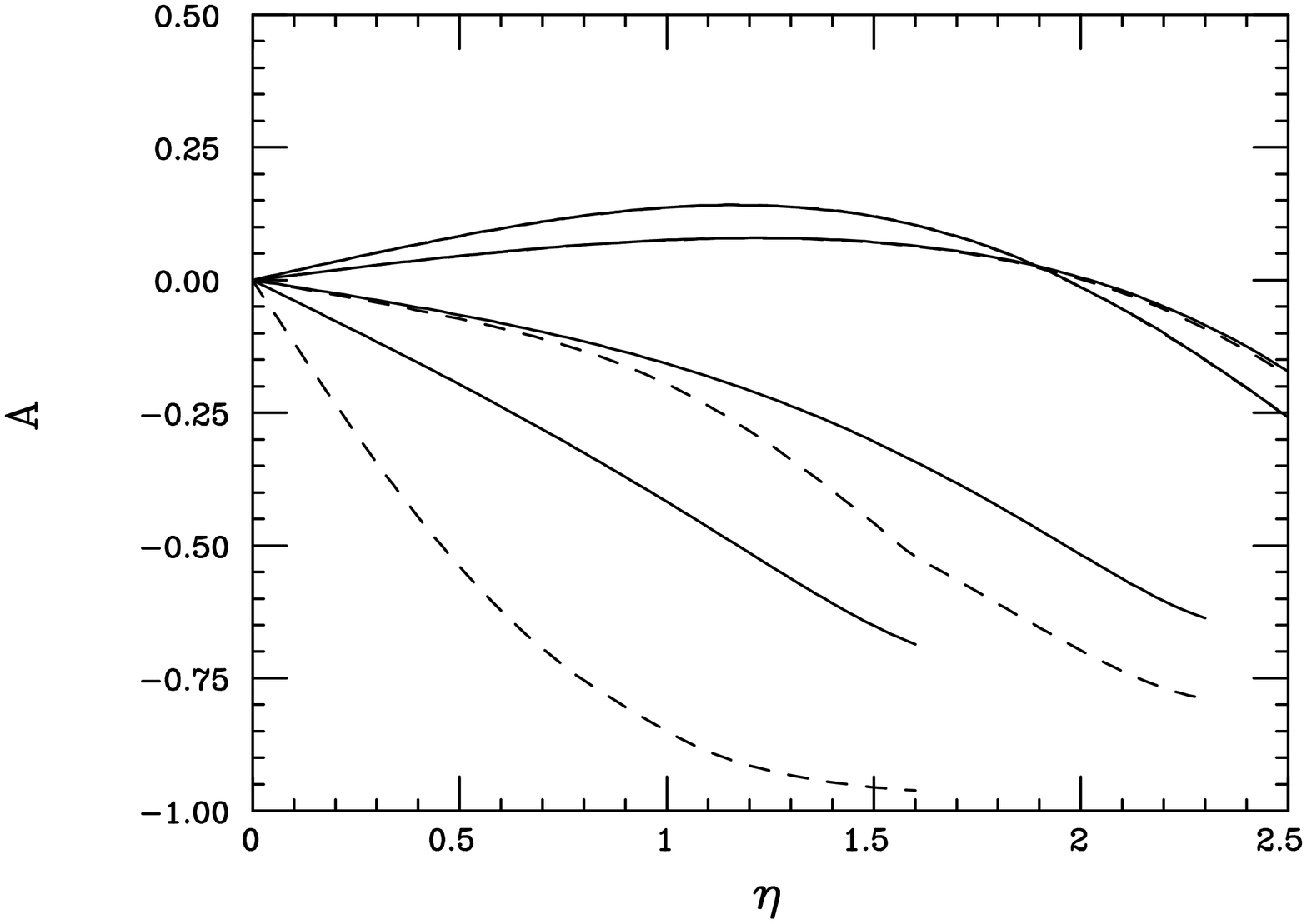,height=4.9cm,width=4.9cm,angle=-90}}
\vspace*{-0.6cm}
\caption{Same as the previous two figures but now for the case of a 800 GeV 
$W'$ with purely left-handed(left panel) or 
purely right-handed(right panel) couplings.}
\label{wprime}
\end{figure}
\vspace*{0.1mm}

Fig.6 shows the corresponding modifications in the leptonic charge asymmetry 
due to an 800 GeV $W'$ with either purely left-handed(LH) or purely 
right-handed(RH) fermionic couplings. Note that the $W'$ with purely RH 
couplings, unlike the LH $W'$, does 
not interfere with the SM amplitude, similar to the case of $\tilde \tau$ 
production. The deviation in the asymmetry due to either type of $W'$ is very 
different than that for a $\tilde \tau$. Here we see that the $W'$ 
substantially increases the magnitude of the asymmetry for 
both coupling types and that RH and LH $W'$ bosons 
are {\it themselves} potentially 
distinguishable by using the data in the $M_T$ bin below but not containing the 
Jacobian peak. 

The $M_T$ bins we have taken in this analysis are rather broad. 
We might expect that if we compress the width of the $M_T$ bin around the $W'$ 
or $\tilde \tau$ Jacobian peak we will increase the purity of the resonant 
contribution and have an even better separation of the two possibilities, at 
the price of reduced statistics. 
(Of course as we narrow this bin we will no longer be able to distinguish 
LH from RH $W'$ bosons 
since this information comes from SM--$W'$ interference.) 
These expectations come to fruition in Fig.7 which shows a more 
direct comparison 
of the lepton charge asymmetries for a $\tilde \tau$ and $W'$ of the same mass 
(800 GeV) and narrowing the width of the $M_T$ bin surrounding the Jacobian 
peak 
to only 300 GeV. Note that the LH and RH $W'$ cases are no longer separable. 
Clearly such measurements will allow the production of $W'$ and $\tilde \tau$ 
to be distinguished.

\vspace*{-0.5cm}
\nn
\begin{figure}[htbp]
\centerline{
\psfig{figure=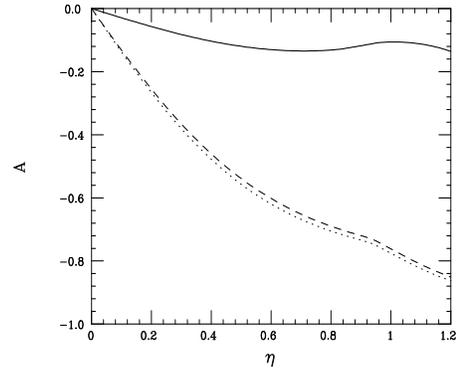,height=6cm,width=7cm,angle=-90}}
\vspace*{-0.6cm}
\caption{Direct comparison of the charge asymmetry induced by a 800 GeV 
$\tilde \tau$(solid) and a LH or RH $W'$(dot and dash) of the same mass 
at the 2 TeV Tevatron. 
For this comparison, a narrow bin in $M_T$ was chosen: $600 < M_T <900$ GeV. 
The Yukawa couplings are as in the earlier figures.}
\label{dis}
\end{figure}
\vspace*{0.1mm}

It is interesting to note that lepton asymmetry deviations can be used to 
probe indirectly for the exchange of $\tilde \tau$ through $R$-parity 
violating couplings. To demonstrate this let us fix the $\tilde \tau$ width to 
mass ratio to be $\Gamma/m=0.004$ and subdivide each of the four $M_T$ bins 
discussed above into rapidity intervals of $\Delta \eta=0.1$. For a given 
$\tilde \tau$ mass we can then ask down to what value of the product of the 
Yukawa couplings, $\lambda \lambda'$, will the asymmetry differ significantly 
from SM expectations. For a fixed mass and integrated luminosity we generate 
Monte Carlo data for various values of the Yukawas and then perform a 
$\chi^2$ analysis to obtain the sensitivity. 
The results of this analysis are shown in Fig.8 and one 
can see that the search reaches obtained in this manner are rather modest.

\vspace*{-0.5cm}
\nn
\begin{figure}[htbp]
\centerline{
\psfig{figure=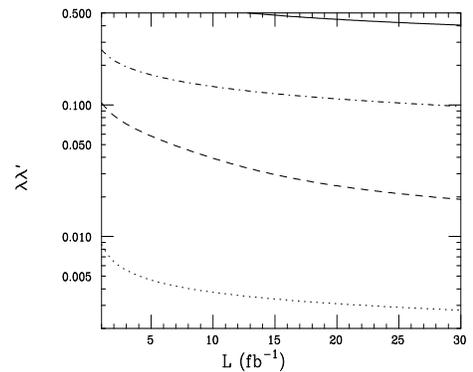,height=6cm,width=7cm,angle=-90}}
\vspace*{-0.6cm}
\caption{Search reach for $\tilde \tau$ exchange as a function of the 2 TeV 
Tevatron integrated luminosity assuming $\Gamma/m=0.004$ for masses of 1500, 
1000, 750 and 250 GeV(from top to bottom).}
\label{disi}
\end{figure}
\vspace*{0.1mm}

\section{The Dijet Channel}

Since $d\bar d$ and/or $u\bar d$ annihilation are responsible for 
producing the slepton resonances, it is obvious that the resonance must 
also decay into these same fermion 
pairs. This means that $\tilde \tau$ or $\tilde \nu$ will decay to dijets and 
may appear as observable peaks above the conventional QCD backgrounds. 
This hope 
will very hard to fulfill at the LHC where QCD backgrounds are expected to be 
severe for searches for narrow resonances which are not strongly produced. At 
the Tevatron, one can be much more optimistic. In fact, searches for such 
narrow dijet resonances have already been performed {\cite {dijetbumps}} at the 
Tevatron  by both CDF and D0 during Run I. Using their 
results and scaling by appropriate factors of beam energy and integrated 
luminosities we may estimate the probable search reaches for CDF and D0 
from Run II . (These estimates conform to the expectations given in 
Ref. {\cite {tev2000}}.) The cross sections themselves are immediately 
calculable in the narrow width approximation in terms of the 
product $Y=(\lambda')^2B_{2j}$, where $\lambda'$ 
is the familiar Yukawa coupling 
and $B_{2j}$ is the dijet branching fraction. The 
results of these calculations are shown in Fig.9. Here we clearly see that for 
values of $Y\sim 0.001-0.01$ or greater, the Tevatron will have a substantial 
mass reach for slepton induced dijet mass bumps during Run II. 
Note that as in the case of 
Drell-Yan, larger cross sections for fixed $Y$ occur in the CC channel than in 
the NC channel due to the larger parton luminosities. Unfortunately, if such 
a bump is observed it will not be straightforward to identify it as a slepton 
resonance.

\vspace*{-0.5cm}
\nn
\begin{figure}[htbp]
\centerline{
\psfig{figure=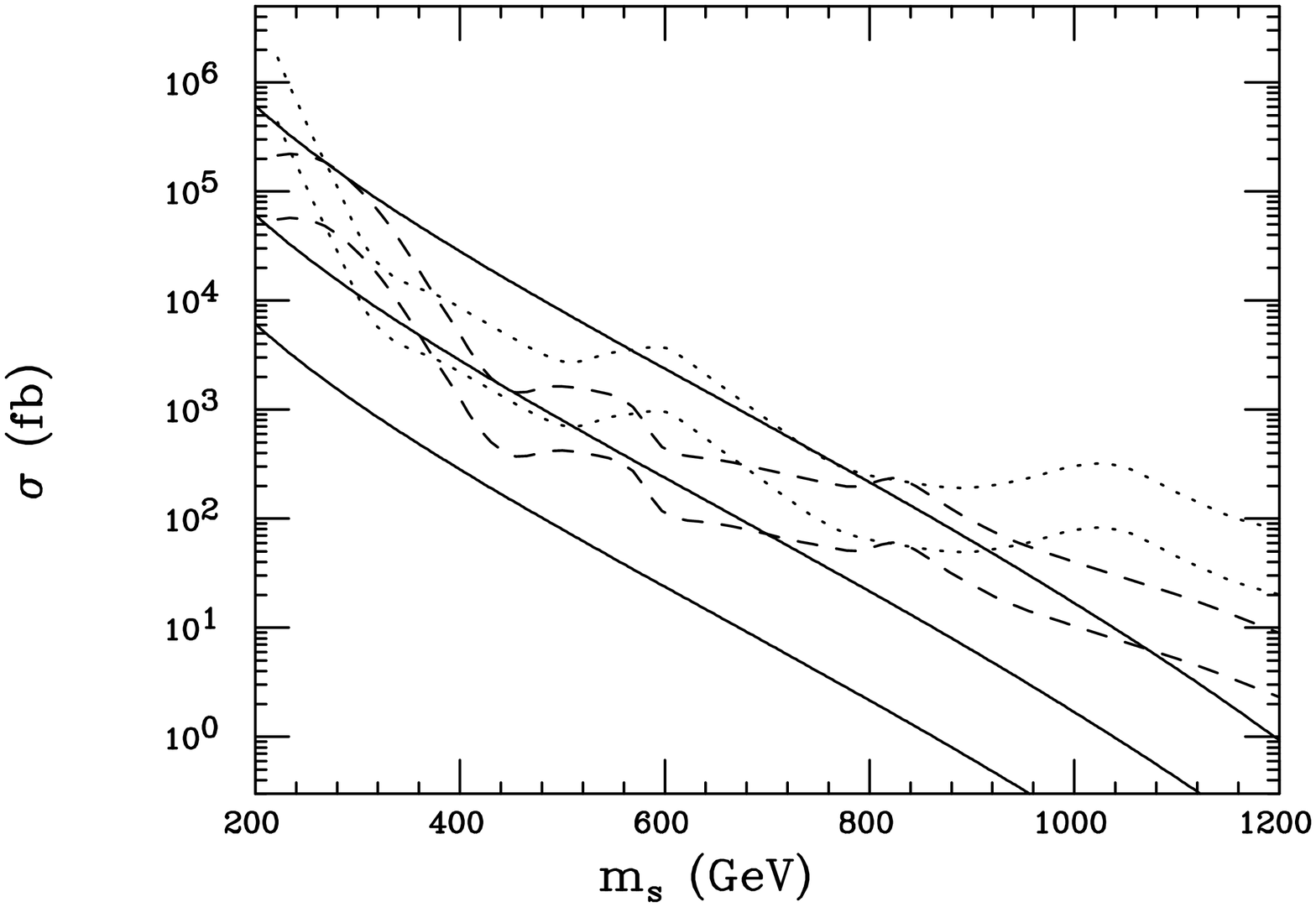,height=4.9cm,width=4.9cm,angle=-90}
\hspace*{-5mm}
\psfig{figure=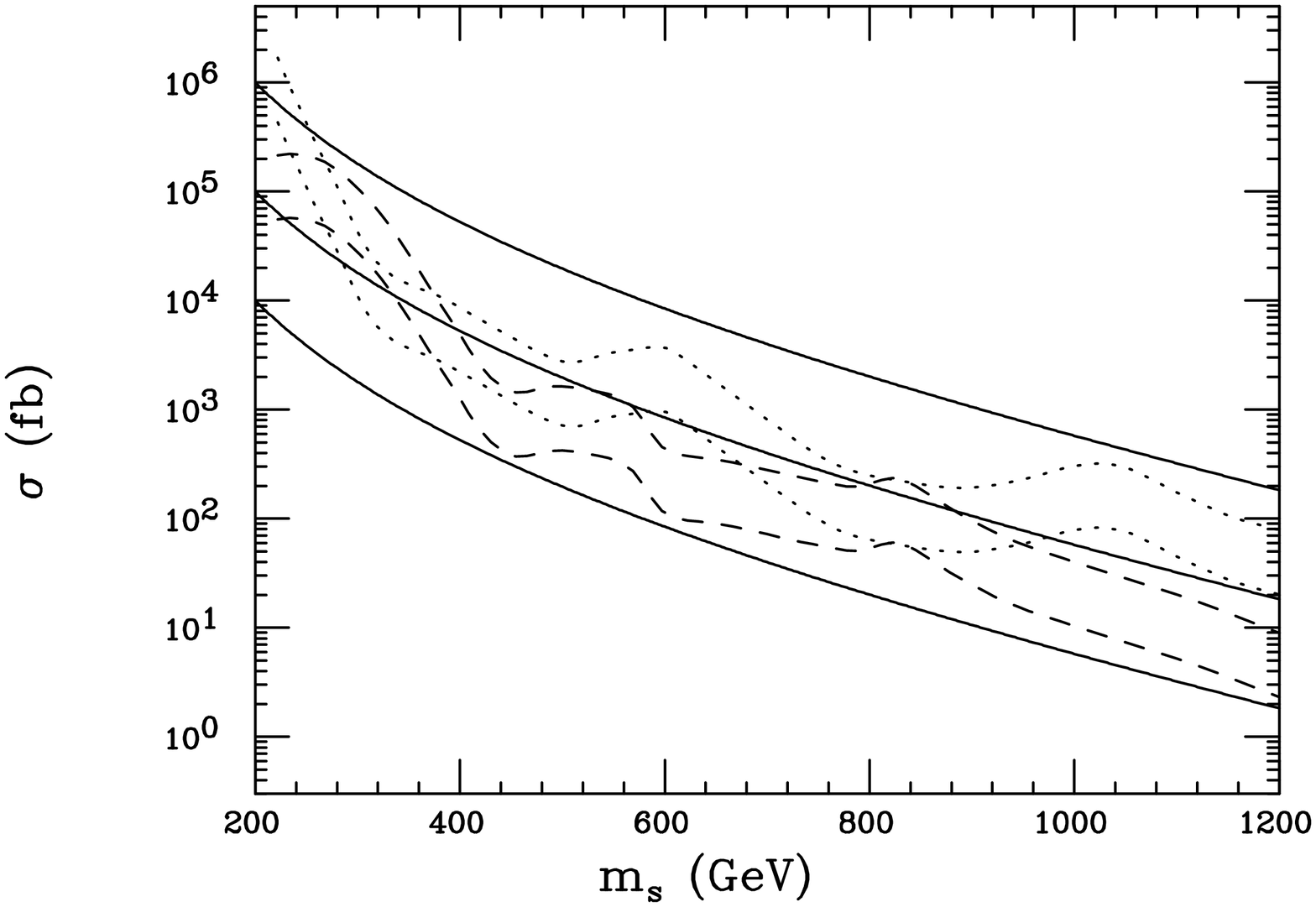,height=4.9cm,width=4.9cm,angle=-90}}
\vspace*{-0.6cm}
\caption{Cross sections for narrow dijet resonances(solid) at the 2 TeV 
Tevatron arising from $\tilde \nu$(left) or $\tilde \tau$(right) production 
in comparison to the 
anticipated search reaches of CDF(dots) and D0(dashes). The upper(lower) 
curve for each experiment assumes an integrated luminosity of 2(30) $fb^{-1}$. 
The three solid curves from top to bottom correspond to slepton resonance 
predictions for $Y$=0.1, 0.01 and 
0.001, respectively, where $Y$ is defined in the text.}
\label{dijets}
\end{figure}
\vspace*{0.1mm}

\section{Conclusion}

As we have seen from the analysis above, resonant $s$-channel production of 
$\tilde \tau$ and/or 
$\tilde \nu$ with their subsequent decay into purely leptonic or dijet 
final states is observable over a wide range of parameters in hadronic 
collisions via $R$-parity violating couplings. We have obtained the 
corresponding search reaches in the slepton mass-$R$-parity violating coupling 
plane for both the Tevatron and LHC. If this signature is observed, 
we have demonstrated that the leptonic angular distributions and the 
lepton charge asymmetry can be successfully used to 
distinguish slepton resonances from those associated with new gauge bosons. 

This process provides a clean and powerful probe of $R$-parity violating 
supersymmetric parameter space.

\end{document}